\documentclass[a4paper]{jpconf}
\usepackage{graphicx}
\begin{document}
\title{Stationary distributions of sums of marginally chaotic variables as renormalization group fixed points}

\author{Miguel A. Fuentes$^{1,2}$  and A. Robledo$^3$}
\address{$^1$ Santa Fe Institute, 1399 Hyde Park Road, Santa Fe, New Mexico 87501, USA}
\address{$^2$ Centro At\'omico Bariloche, Instituto Balseiro and CONICET, 8400 Bariloche, Argentina}
\address{$^3$ Instituto de F\'{\i}sica, Universidad Nacional Aut\'onoma de M\'exico, Apartado Postal 20-364, M\'exico 01000 DF, Mexico}

\ead{fuentesm@santafe.edu}

\begin{abstract}
We determine the limit distributions of sums of deterministic chaotic variables in unimodal maps assisted by a novel renormalization group (RG) framework associated to the operation of increment of summands and rescaling. In this framework the difference in control parameter from its value at the transition to chaos is the only relevant variable, the trivial fixed point is the Gaussian distribution and a nontrivial fixed point is a multifractal distribution with features similar to those of the Feigenbaum attractor. The crossover between the two fixed points is discussed and the flow toward the trivial fixed point is seen to consist of a sequence of chaotic band mergers.
\end{abstract}

\section{Introduction}
~

The hegemony of the Central Limit Theorem \cite{vankampen1, khinchin1} for 
sums of deterministic variables generated by a number chaotic mappings have 
for some time been observed and also mathematically proved  \cite{kaminska1}. Since 
the mixing properties of chaotic trajectories yield variables indistinguishable 
to independent random variables, it is of interest to study nonmixing systems 
such as mappings at the transition from regular to chaotic behavior. Recent 
\cite{tsallis}-\cite{grassberger1} numerical explorations of time averages of iterates at the 
period-doubling transition to chaos \cite{schuster1}  have been presented and 
interpreted as possible evidence for a novel type of stationary distribution.

The dynamics toward and at the Feigenbaum attractor is now known in much
detail \cite{robledo1, robledo2}, therefore, it appears feasible to analyze
also the properties of sums of iterate positions for this classic nonlinear
system with the same kind of analytic reasoning and numerical thoroughness.
Here we present the results for sums of chronological positions of
trajectories associated to\ quadratic unimodal maps. We consider the case of
the sum of positions of trajectories inside the Feigenbaum attractor as well
as those within the chaotic $2^{n}$-band attractors obtained when the
control parameter is shifted to values larger than that at the transition to
chaos. From the information obtained we draw conclusions on the properties
of the stationary distributions for these sums of variables. Our results,
that reveal a multifractal stationary distribution that mirrors the features
of the Feigenbaum attractor, can be easily extended to other critical
attractor universality classes and other routes to chaos.

The overall picture we obtain is effectively described within the framework
of the renormalization group (RG) approach for systems with scale invariant
states or attractors. Firstly, the RG transformation for the distribution of
a sum of variables is naturally given by the change due to the increment of
summands followed by a suitable restoring operation. Second, the limit
distributions can be identified as fixed points reached according to whether
the acting relevant variables are set to zero or not. Lastly, the
universality class of the non-trivial fixed-point distribution can be
assessed in terms of the existing set of irrelevant variables. 

As it is well
known \cite{schuster1} a few decades ago the RG approach was successfully
applied to the period-doubling route to chaos displayed by unimodal maps. In
that case the RG transformation is functional composition of the mapping and
its effect re-enacts the growth of the period doubling cascade. In our case
the RG transformation is the growth and adjustment of the sum of positions
and its effect is instead to go over again the merging of bands in the
chaotic region.

Specifically, we consider the Feigenbaum map $g(x)$, obtained from the fixed
point equation $g(x)=\alpha g(g(x/\alpha ))$ with $g(0)=1$ and $g^{\prime
}(0)=0$, and where $\alpha =-2.50290...$ is one of Feigenbaum's universal
constants \cite{schuster1}. For expediency we shall from now on denote the
absolute value $\left\vert \alpha \right\vert $ by $\alpha $. Numerically,
the properties of $g(x)$ can be conveniently obtained from the logistic map $%
f_{\mu ,2}(x)=1-\mu x^{2},\;-1\leq x\leq 1$, with $\mu =$ $\mu _{\infty
}=1.401155189092..$. The dynamics associated to the Feigenbaum map is
determined by its multifractal attractor. For a recent detailed
description of these properties see \cite{robledo1, robledo2}. For values of 
$\mu >$ $\mu _{\infty }$ we employ a well-known scaling relation supported
by numerical results.

Initially we present properties of the sum of the absolute values $%
\left\vert x_{t}\right\vert $ of positions $x_{t}=f_{\mu _{\infty
},2}(x_{t-1}),\;t=1,2,3,...$, as a function of total time $N$ visited by the
trajectory with initial position $x_{0}=0$, and obtain a patterned linear
growth with $N$. We analyze this intricate fluctuating pattern, confined
within a band of finite width, by eliminating the overall linear increment
and find that the resulting stationary arrangement exhibits features
inherited from the multifractal structure of the attractor. We derive an
analytical expression for the sum that corroborates the numerical results
and provide an understanding of its properties. Next, we consider the
straight sum of $x_{t}$, where the signs taken by positions lessen the
growth of its value as $N$ increases and the results are consistently
similar to those for the sum of $\left\vert x_{t}\right\vert $, i.e. linear
growth of a fixed-width band within which the sum displays a fluctuating
arrangement. Further details for the sum of $x_{t}$ are not included because
of repetitiveness. Then, we show numerical results for
the sum of iterated positions obtained when the control parameter is shifted
into the region of chaotic bands. In all of these cases the distributions
evolve after a characteristic crossover towards a Gaussian form. Finally, we
rationalize our results in terms of an RG framework in which the action of
the Central Limit Theorem plays a fundamental role.

\begin{figure}[tbp]
\centering \includegraphics[width=11cm,angle=0]{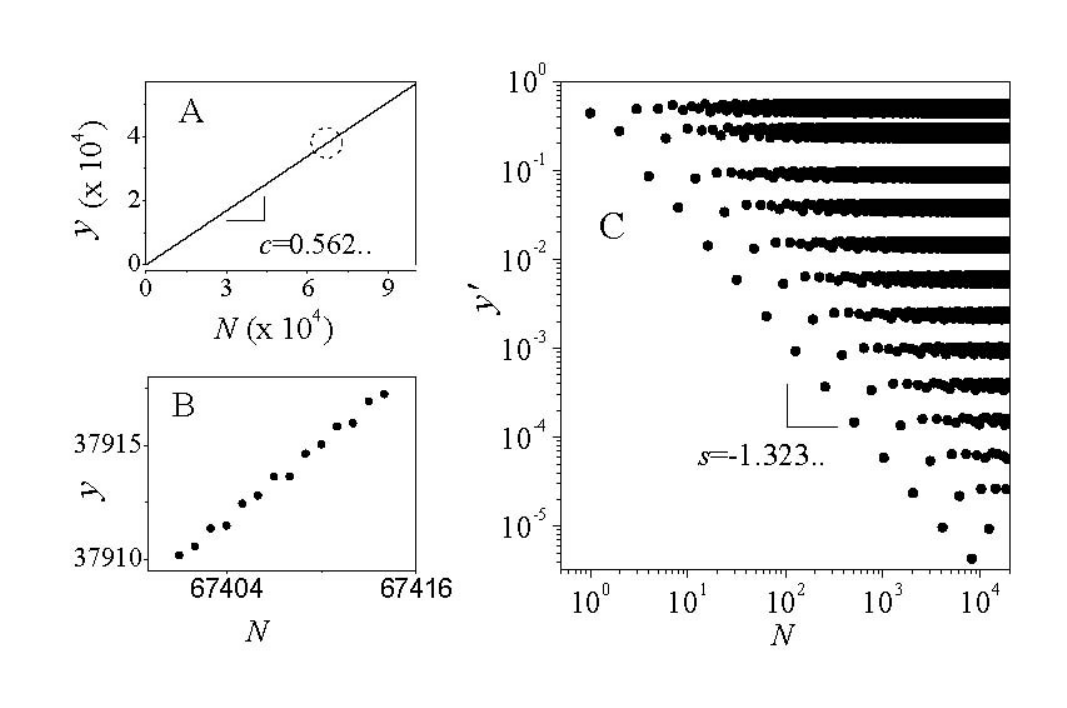}\newline
\caption{A) Sum of absolute values of visited points $x_{t}$, $t=0,...,N$, of
the Feigenbaum's attractor with initial condition $x_{0}=0$. B) A
closer look of the path of the sum (see dotted circle in A), for values of $N$ around 67410. C) Centered sum $y^{\prime }(N)$ in logarithmic scales. See
text.}
\label{f1}
\end{figure}

~

The starting point of our study is evaluation of 
\begin{equation}
y_{\mu }(N)\equiv \sum\limits_{t=1}^{N}\left\vert x_{t}\right\vert ,
\label{sumabs1}
\end{equation}%
with $\mu =\mu _{\infty }$ and with $x_{0}=0$. Fig. 1A shows the result,
where it can be observed that the values recorded, besides a repeating
fluctuating pattern within a narrow band, increase linearly on the whole.
The measured slope of the linear growth is $c=0.56245...$ Fig. 1B shows an
enlargement of the band, where some detail of the complex pattern of values
of $y_{\mu _{\infty }}(N)$ is observed. A stationary view of the mentioned
pattern is shown in Fig. 1C, where we plot%
\begin{equation}
y_{\mu _{\infty }}^{\prime }(N)\equiv \sum\limits_{t=1}^{N}\left( \left\vert
x_{t}\right\vert -c\right) ,  \label{sumabs2}
\end{equation}%
in logarithmic scales. There, we observe that the values of $y_{\mu _{\infty
}}^{\prime }(N)$ fall within horizontal bands interspersed by gaps,
revealing a fractal or multifractal set layout. The top (zeroth) band
contains $y_{\mu _{\infty }}^{\prime }$ for all the odd values of $N$, the
1st band next to the top band contains $y_{\mu _{\infty }}^{\prime }$ for
the even values of $N$ of the form $N=2+4m$, $m=0,1,2,...$ The 2nd band next
to the top band contains $y_{\mu _{\infty }}^{\prime }(N)$ for $%
N=2^{2}+2^{3}m$, $m=0,1,2,...$, and so on. In general, the $k$-th band next
to the top band contains $y_{\mu _{\infty }}^{\prime }(2^{k}+2^{k+1}m)$, $%
m=0,1,2,...$ Another important feature in this figure is that the $%
y_{\mu _{\infty }}^{\prime }(N)$ for subsequences of $N$ each of the form $%
N=(2l+1)2^{k}$, $k=0,1,2,...$, with $l$ fixed at a given value of $%
l=0,1,2,...$, appear aligned with a uniform slope $s=-1.323...$ The parallel lines formed by these subsequences imply the power law $%
y_{\mu _{\infty }}^{\prime }(N)\sim N^{s}$ for $N$ belonging to such a
subsequence.

It is known \cite{robledo1, robledo3} that these two characteristics of $%
y_{\mu _{\infty }}^{\prime }(N)$ are also present in the layout of the
absolute value of the individual positions $\left\vert x_{t}\right\vert $, $%
t=1,2,3,...$ of the trajectory initiated at $x_{0}=0$; and this layout
corresponds to the multifractal geometric configuration of the points of the
Feigenbaum's attractor, see Fig. 1 in \cite{robledo3}. In this case, the
horizontal bands of positions separated by equally-sized gaps are related to
the period-doubling `diameters' \cite{schuster1} set construction of the
multifractal \cite{robledo2}. The identical slope shown in the logarithmic
scales by all the position subsequences $\left\vert x_{t}\right\vert $, $%
t=(2l+1)2^{k}$, $k=0,1,2,...$, each formed by a fixed value of $l=0,1,2,...$%
, implies the power law $\left\vert x_{t}\right\vert \sim t^{s}$, $s=-\ln
\alpha /\ln 2=-1.3236...$, as the $\left\vert x_{t}\right\vert $ can be
expressed as $\left\vert x_{t}\right\vert \simeq \left\vert
x_{2l+1}\right\vert \ \alpha ^{-k},t=(2l+1)2^{k}$, $k=0,1,2,...$, or,
equivalently, $\left\vert x_{t}\right\vert \sim t^{s}$. Notice that the
index $k$ also labels the order of the bands from top to bottom. The power
law behavior involving the universal constant $\alpha $ of the subsequence
positions reflect the approach of points in the attractor toward its most
sparse region at $x=0$ from its most compact region, as the positions at odd
times $\left\vert x_{2l+1}\right\vert =x_{2l+1}$, those in the top band,
correspond to the densest region of the set.

Having uncovered the through manifestation of the multifractal structure of
the attractor into the sum $y_{\mu _{\infty }}^{\prime }(N)$ we proceed to
derive this property and corroborate the numerical evidence. Consider Eq. (%
\ref{sumabs1}) with $N=2^{k}$, $k=0,1,2,...$, the special case $l=0$\ in the
discussion above. Then the numbers of terms $\left\vert x_{t}\right\vert $
per band in $y_{\mu _{\infty }}(2^{k})$ are: $2^{k-1}$ in the top band\ ($%
j=0 $), $2^{k-2}$ in the next band\ ($j=1$),..., $2^{0}$ in the ($k-1$)-th
band, plus an additional position in the $k$-th band. If we introduce the
average of the positions on the top band%
\begin{equation}
\left\langle a\right\rangle \equiv 2^{-(k-1)}\sum_{j=0}^{2^{k-1}}x_{2j+1},
\label{ave1}
\end{equation}%
the sum $y_{\mu _{\infty }}(2^{k})$ can be written as%
\begin{equation}
y_{\mu _{\infty }}(2^{k})=\left\langle a\right\rangle
2^{k-1}\sum_{j=0}^{k-2}(2\alpha )^{-j}+\alpha ^{-(k-1)}+\alpha ^{-k}.
\label{sumabs3}
\end{equation}%
Doing the geometric sum \ above and expressing the result as $y_{\mu
_{\infty }}(2^{k})=c2^{k}+d\alpha ^{-k}$, we have%
\begin{equation}
c=\frac{\left \langle a \right \rangle \alpha}{2\alpha -1}, ~~~d=\left( 1-\frac{\left\langle a\right\rangle \ 2\alpha }{2\alpha -1}\right)
\ \alpha +1\ .  \label{slopeandshift1}
\end{equation}%
Evaluation of Eq. (\ref{ave1}) yields to $\left\langle a\right\rangle
=0.8999...$, and from this we obtain $c=0.56227...$ and $d=0.68826...$ We
therefore find that the value of the slope $c$ in Fig. 1A is properly
reproduced by our calculation. Also, since $\ln \left[ y_{\mu _{\infty
}}(2^{k})-c2^{k}\right] =\ln d-k\ln \alpha $, or, equivalently, $\ln y_{\mu
_{\infty }}^{\prime }(N)=\ln d-N\ln \alpha /\ln 2$,$\;N=2^{k}$,$%
\;k=0,1,2,... $, we corroborate that the value of the slope $s$ in inset of
Fig. 1C is indeed given by $s=-\ln \alpha /\ln 2=1.3236...$ (We have made
use of the identity $\alpha ^{-k}=N^{-\ln \alpha /\ln 2}$,$\;N=2^{k}$,$%
\;k=0,1,2,...$).

We note that the sum of $x_{t}$ from $t=0$ to $N=2^{k}$, i.e. considering
the signs taken by positions, can be immediately obtained from the above by
replacing $\alpha ^{-j}$ by $(-1)^{j}\alpha ^{-j}$ as the $x_{t}$ of
different signs of the trajectory starting at $x_{0}=0$ fall into separate
alternating bands (described above and shown in Fig. 1 of \cite{robledo3}).
In short, $x_{t}\simeq (-1)^{j}x_{2l+1}\ \alpha ^{-j}$, $t=(2l+1)2^{k}$, $%
k=0,1,2,...$ As stated, our numerical and analytical results are in
agreement also in this case.
\begin{figure}[tbp]

\centering \includegraphics[width=11cm,angle=0]{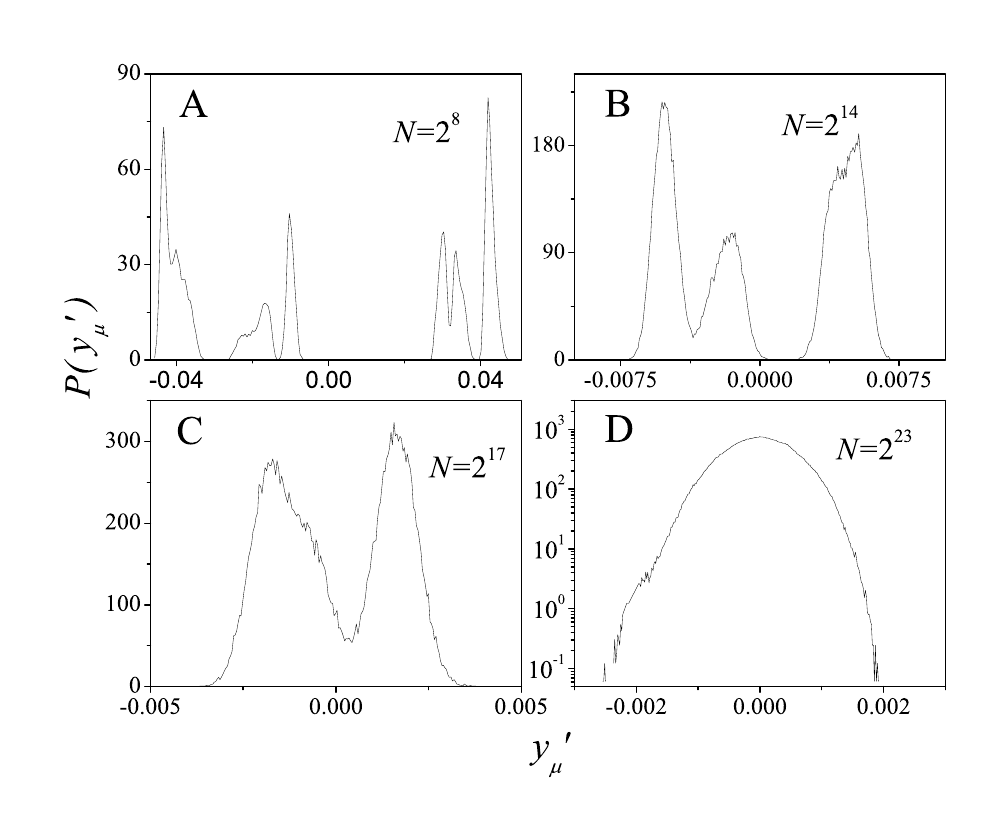}\newline
\caption{Distributions for the sums of $\left\vert x_{t}\right\vert $, $t=0,...,N$, of
an ensemble of trajectories with initial conditions within the $2^{3}$-band attractor at $\Delta \mu=0.0028448109$. The number of summands $N$ are indicated in each panel. See text.}
\label{f1}
\end{figure}

We turn now to study the sum of positions of trajectories when $\Delta \mu
\equiv \mu -$ $\mu _{\infty }>0$. We recall that in this case the attractors
are made up of $2^{K}$, $K=1,2,3,...$, bands and that their trajectories
consist of an interband periodic motion of period $2^{K}$ and an intraband
chaotic motion. We evaluated numerically the sums $y_{\mu }(N)$ for an ensemble
of  initial conditions $x_{0}$ uniformly distributed only within the chaotic
bands, for different values of $\Delta \mu $;  $y_{\mu }^{\prime }(N)$ was then obtained
similarly to Eq. (\ref{sumabs2}) by substracting the average $\left\langle
y_{\mu }(N)\right\rangle _{x_{0}}$ and rescaling with a factor $N^{-1/2}$. The panels
in Fig. 2 show the evolution of the distributions for increasing number of
summands $N$ for a value of $\Delta \mu $ (chosen for visual clarity) when
the attractor consists of $2^{3}$ chaotic bands. Initially the distributions
are multimodal with disconnected domains, but as $N$ increases we observe
merging of bands and development of a single-domain bell-shaped distribution
that as $N\longrightarrow \infty $ converges in all cases to a Gaussian
distribution.

These numerical results can be understood as follows. We recollect \cite%
{schuster1} that the relationship between the number $2^{K}$, $K\gg 1$, of
bands of a chaotic attractor and the control parameter distance $\Delta \mu $
at which it is located is given by $2^{K}\sim \Delta \mu ^{-\kappa }$, $%
\kappa =\ln 2/\ln \delta _{F}$, where $\delta _{F}=$ $0.46692...$ is the
universal constant that measures both the rate of convergence of the values
of $\mu $ at period doublings or at band splittings to $\mu _{\infty }$. For $\Delta \mu $ small and fixed, the sum of sequential
positions of the trajectory initiated at $x_{0}=0$, Eq. (\ref{sumabs1}),
exhibits two different growth regimes as the total time $N$ increases. In
the first one, when $N\ll 2^{K}$, the difference in value $\delta
x_{t}\equiv x_{t}(\mu )-x_{t}(\mu _{\infty })$ between the positions at time 
$t$ for $\mu $ and $\mu _{\infty }$ do not affect qualitatively the
multifractal structure of the sum at $\mu _{\infty }$, nor its associated
distribution. This is because the fine structure of the Feigenbaum attractor
is not suppressed by the fluctuations $\delta x_{t}$, as these contribute to
the sum individually during the first cycle of the interband periodic
motion. The discrete multi-scale nature of the distribution for $\mu
_{\infty }$ is preserved when the interband motion governs the sum $y_{\mu
}(N)$. In the second regime, when $N\gg 2^{K}$, the situation is opposite,
after many interband cycles the fluctuations $\delta x_{t}$ add up in the
sum and progressively wipe up the fine structure of the Feigenbaum
attractor, leading to merging of bands and to the dominance of the
fluctuating intraband motion. Ultimately, as $N\longrightarrow \infty $ the
evolution of the distribution is similar to the action of the Central Limit
Theorem and leads to a Gaussian stationary end result. It is also evident
that as $\Delta \mu $ increases the first regime is shortened at the expense
of the second, whereas when $\Delta \mu \longrightarrow 0$ the converse is
the case. Therefore there exists an unambiguous $\Delta \mu $-dependent
crossover behavior between the two radically different types of stationary
distributions. This crossover is set out when the $\delta x_{t}$
fluctuations begin removing the band structure in $%
y_{\mu }^{\prime }(N)$ when $\Delta \mu $ is small and ends when these
fluctuations have broadened and merged all the chaotic bands and $y_{\mu
}^{\prime }(N)$ forms a single continuous interval. When $\mu =\mu _{\infty
} $ this process never takes place.

We are in a position now to put together the numerical and analytical
information presented above into the general framework of the RG approach.
As known, this method was designed to characterize families of systems
containing amongst their many individual states (or in this case attractors)
a few exceptional ones with scale invariant properties and common to all
systems in the family. We recall \cite{fisher1} that in the language of a
minimal RG scheme there are two fixed points, each of which can be reached
by the repeated application of a suitable transformation of the system's
main defining property. One of the fixed points, is termed trivial and is
reached via the RG transformation for almost all initial settings. i.e. for
all systems in the family when at least one of a small set of variables,
named relevant variables, is nonzero. To reach the other fixed point, termed
nontrivial, it is necessary that the relevant variables are all set to zero,
and this implies a severely restricted set of initial settings that ensure
such critical RG paths. The nontrivial fixed point embodies the scale
invariant properties of the exceptional state that occurs in each system in
the family and defines a universality class, while the differences amongst
the individual systems are distinguished through a large set of so-called
irrelevant variables. The variables in the latter set gradually vanish as
the RG transformation is applied to a system that evolves toward the
nontrivial fixed point. Further, when any system in the family is given a
nonzero but sufficiently small value to (one or more of) the relevant
variables, the RG transformation converts behavior similar to that of the
nontrivial fixed point into that resembling the trivial fixed point through
a well-defined crossover phenomenon. The recognition of the RG framework in
the properties of the sums of positions of trajectories in unimodal maps and
their associated distributions is straightforward. It can be concluded right
away that in this problem (as defined here) there is only one relevant
variable, the control parameter difference $\Delta \mu $. There is an
infinite number of irrelevant variables, those that specify the differences
between all possible unimodal maps (with quadratic maximum) and the
Feigenbaum map $g(x)$. There are two fixed-point distributions, the trivial
continuum-space Gaussian distribution and the nontrivial discrete-space
multifractal distribution (as observed in Fig. 1C). As explained above,
there is a distinct crossover link between the two fixed-point
distributions. The RG transformation consists of the increment of one or
more summands in the sum (\ref{sumabs1}) followed by centering like in Eq. (%
\ref{sumabs2}). The effect of the transformation in the distribution of the
sum is then recorded. For sums of independent variables the transformation
is equivalent to the convolution of distributions. Our results correspond to
the dynamics inside the atractors, however, if the
interest lies in considering only the stationary distribution of sums that
do not contain the transient behavior of trajectories in their way to the
attractor \cite{tsallis} our results are expected to give the correct
answers for this case.

In summary, we have found that the stationary distribution of the sum of
iterate positions within the Feigenbaum attractor has a multifractal
structure stamped by that of the initial multifractal set, while that
involving sums of positions within the attractors composed of $2^{K}$
chaotic bands is the Gaussian distribution. We have also shown that the
entire problem can be couched in the language of the RG formalism in a way
that makes clear the identification of the existing stationary distributions
and the manner in which they are reached. These basic features suggest a
degree of universality, limited to the critical attractor under
consideration, in the properties of sums of deterministic variables at the
transitions to chaos. Namely, the sums of positions of memory-retaining
trajectories evolving under a vanishing Lyapunov exponent appear to preserve
the particular features of the multifractal critical attractor under
examination. Thus we expect that varying the degree of nonlinearity of a
unimodal map would affect the scaling properties of time averages of
trajectory positions at the period doubling transition to chaos, or
alternatively, that the consideration of a different route to chaos, such as
the quasiperiodic route, would lead to different scaling properties of
comparable time averages. For instance, the known dependence of the
universal constant $\alpha $ on the degree of nonlinearity $z$ of a unimodal
map would show as a $z$-dependent exponent $s=\ln \alpha /\ln 2$ that
controls the scale invariant property of the sum of trajectory positions
with $x_{0}=0$ (shown in Fig. 1C ).

We analyzed the nature and the conditions under which a stationary distribution with universal properties (in the Renormalization Group sense) occurs for sums of deterministic variables at the transition between regular and chaotic behavior, such as those studied here for dynamics at zero Lyapunov exponent. The nonexistence of fluctuations in such critical attractors implies a distribution of the sums of these variables strictly defined on a discrete multifractal set and therefore different from known (Gaussian or otherwise) continuum-space limit distributions for real number random variables.

\ack We appreciate partial financial support by
DGAPA-UNAM and CONACYT (Mexican agencies). AR is grateful for hospitality
received at the SFI.

\end{document}